# Effect of Diffused Hydrogen on the Conductance of Fe/Mgo/Fe Magnetic Tunnel Junctions: Atomistic Simulation

Ankit Kumar Verma[1] and Bahniman Ghosh[2,3]

[1]Electronics and Communication Engineering Department, National Institute of Technology, Hamirpur, Himachal Pradesh, India

[2]Department of Electrical Engineering, Indian Institute of Technology Kanpur, Kanpur 208016, India.

[3]Microelectronics Research Center, 10100 Burnet Road, Bldg. 160, University of Texas at Austin, Austin, TX, 78758, USA. (bghosh@utexas.edu)

*Abstract-*
In this work we have analysed the deterimental effect on conductance caused by diffusion of hydrogen atoms in interstitial voids during fabrication process of magnesium oxide barrier of an Fe/MgO/Fe magnetic tunnel junction, using first principle calculations. This diffusion of hydrogen atoms in interstitial voids in barrier disturbs a certain kind of symmetry possessed by magnesium oxide often termed as $\Delta_1$ state symmetry. Distortion in $\Delta_1$ state symmetry tempers the condutance through the magnetic tunnel junction specially in parallel configuration, which in turn reduces the Tunnelling Magneto-Resistance (TMR) to large extent. Improved methods of fabrication can be used to counter such kind of problems which affect device operation.

**INTRODUCTION**

In this era of devloping technology spintronics became a field of interest for many researchers beacuse of its key features such as non volatile processing devices, very low power dissipiation in idle state and many others. In spintronics transport carried out through the device depends on electron spin orientation. Using principles of spintronics Magnetic Random Access Memories (MRAM's) have been designed with basic constituent as Magnetic Tunnel Junctions (MTJ's)[1-5]. Unlike other memories MRAM's never require any kind of refresh and spin torque transfer technique is used for switching which in turn reduces the power dissipiation to a large extent.

Structurally a magnetic tunnel junction consists of two electrodes with a tunneling barrier, preferably an oxide material, in between. One of the electrodes is termed as free layer and other one as fixed layer. Data is stored as relative magnetization orientation of these two layers, parallel orientation as one state and anti-parallel orientation as other state. Tunnelling Magneto-Resistance is proportional to difference between individual resistances offered in transport during parallel and anti-parallel magnetization orientation of layers[4-11]. Hydrogen is a very small atom and it can get trapped into an interstial void of magnesium oxide bulk. Magnesium Oxide interfacing with a metal shows more capability for hydrogen diffusion in comparison to pure magnesium layer. After trapping into the void it tends to make a bond with an oxygen atom[10-14].

Magnesium Oxide possesses $\Delta_1$ state symmetry which has compatibility with a specific set of atomic orbital character sharing. As discussed above hydrogen atom doped in interstitial void tends to make a bond with oxygen which dilutes orbital character sharing between magnesium and oxygen resulting in distortion of $\Delta_1$ state symmetry and in turn tunnelling magneto-resistance of magnetic tunnel junction is reduced to a much lower value[9-13]. This lowering of TMR is because of reduction in conductance through conduction channels compatible with $\Delta_1$ state symmetry.

**MODELING AND SIMULATIONS**

Structure of magnetic tunnel junction has been modelled using Atomistix Tool Kit (ATK) embedded with first principle calculator using Density Functional Theoy (DFT). An ideal Fe/MgO/Fe MTJ has been optimized using DFT with adaption of exhange correlation as Spin dependent Gerenalized Gradient Approximation (SGGA), grid mesh cut-off energy of 300 Rydberg and with a force tolerance of 0.05 eV/Å for super cell having K-mesh of 6×6×2. Fig.1 shows the optimized structure of Fe/MgO/Fe MTJ without any doping. In this basis set used for Fe atom electron wave is Single Zeta Polarized ($\zeta$) while for Mg and O atom electron wave is Double Zeta ($\zeta$)[15-19].

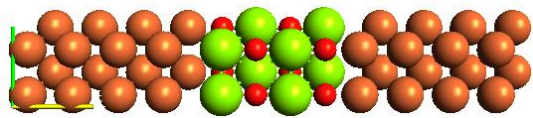

**Fig1. Optimized structure of Fe/MgO/Fe MTJ**

Fig.2 shows optimized structure of same MTJ with doping of Hydrogen atoms in interstitial voids in tunnelling barrier. It is clearly evident from Fig.2 that hydrogen atoms shown by blue colour are trapped in the voids and have a bond with nearest oxygen atom. In this case of optimiztion all the parameters are kept same and one additional parameter basis set for hydrogen atom electron wave is kept as Double Zeta ($\zeta$).

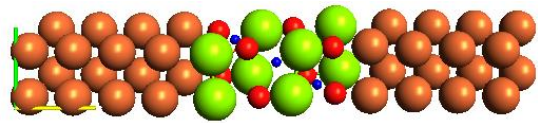

**Fig2. Optimized structure of Hydrogen Diffused Fe/MgO/Fe MTJ**

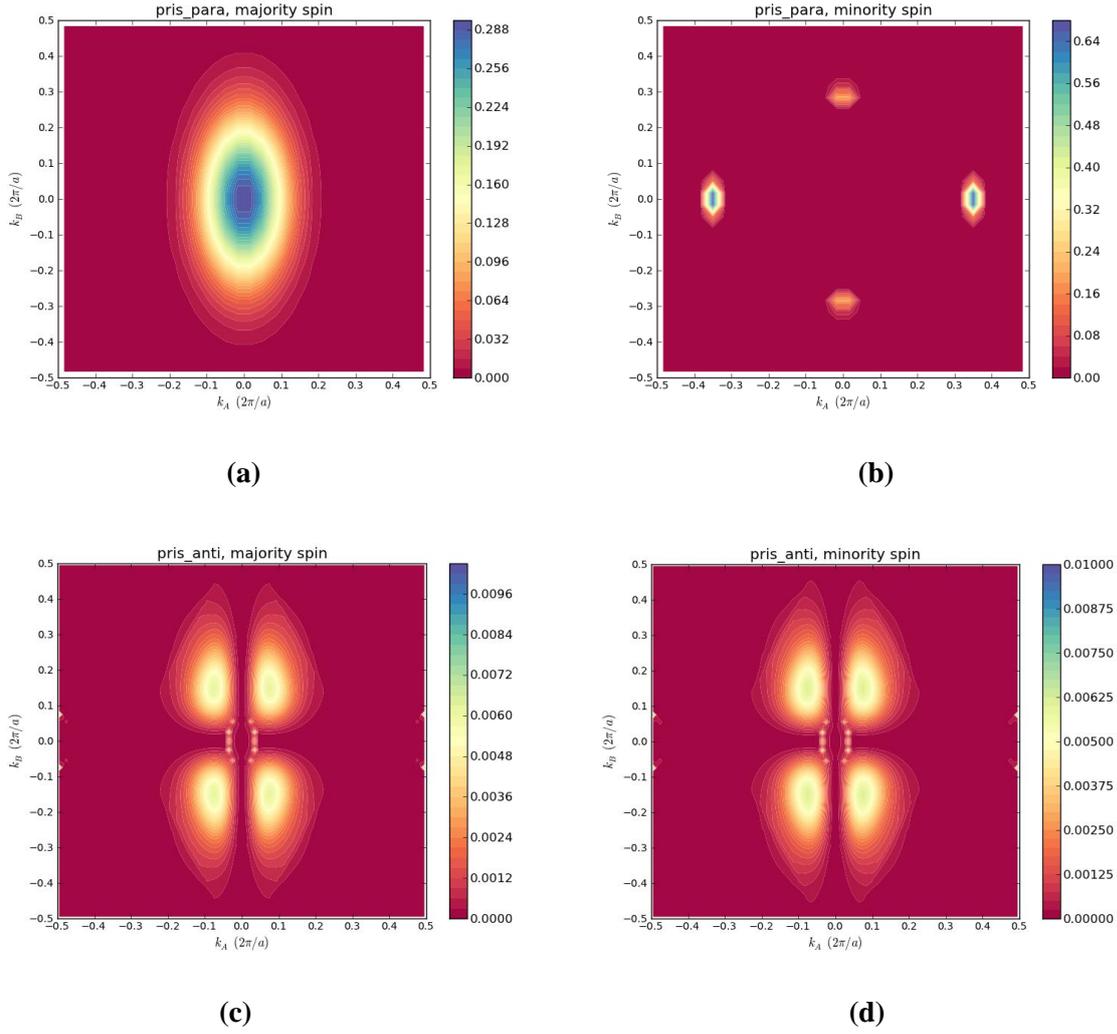

Fig.3: Spin dependent transmission spectrum with a function of in plane $k_{||} = (k_A, k_B)$ wave vector of Fe/MgO/Fe MTJ (a) Majority Spin in Parallel Orientation (b) Minority Spin in Parallel Orientation (c) Majority Spin in anti-parallel Orientation (b) Minority Spin in anti-parallel Orientation

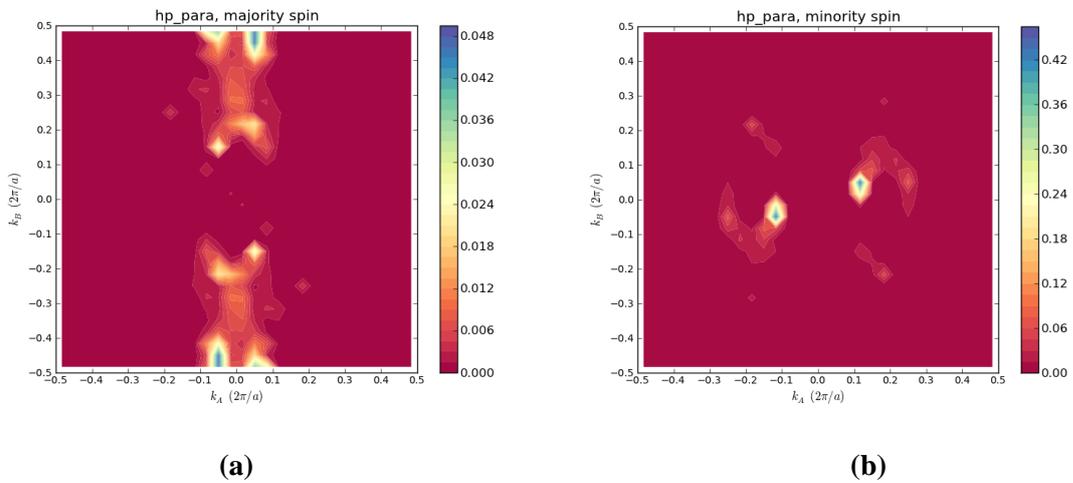

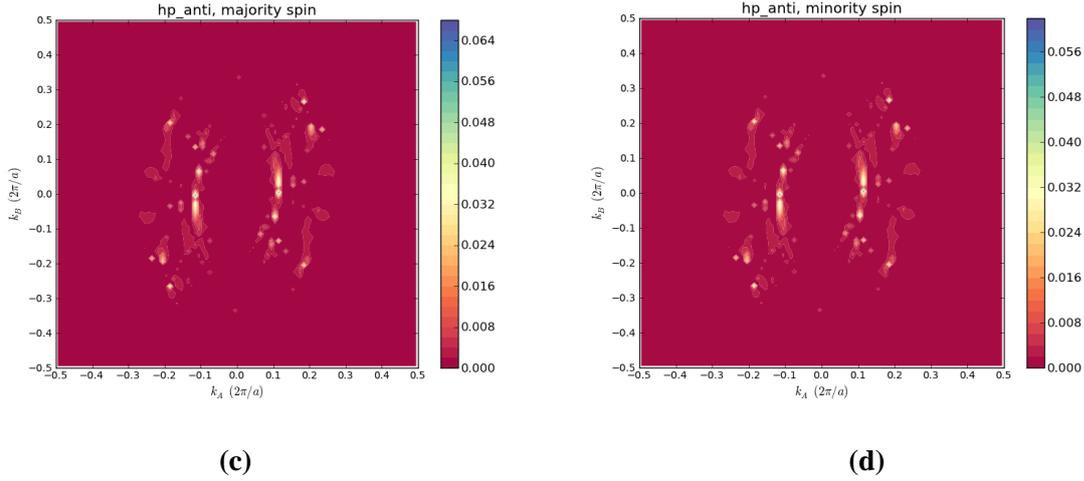

(c)    (d)

**Fig.4:** Spin dependent transmission spectrum with a function of in plane $k_{\parallel} = (k_A, k_B)$ wave vector of Hydrogen Diffused in Tunnelling Barrier of Fe/MgO/Fe MTJ (a) Majority Spin in Parallel Orientation (b) Minority Spin in Parallel Orientation (c) Majority Spin in anti-parallel Orientation (b) Minority Spin in anti-parallel Orientation

TMR calculations have been carried out using Monkhorst pack K mesh of 6×6×100 and with the same parameters K dependent transmission spectrum are calculated and contour plot of this also has been plotted[17-21]. Fig.3 shows contour plot of Spin dependent transmission spectrum with a function of in plane $k_{\parallel} = (k_A, k_B)$ wave vector for ideal Fe/MgO/Fe MTJ while Fig.4 shows the same for MTJ with hydrogen diffused in interstitial void of its tunnelling barrier. Brighter spots shown in Fig.3 and Fig.4 depict the areas having higher condutance value in $k_{\parallel}$ direction in perpendicular to current flow and in parallel with electrode barrier interface.

**RESULTS AND DISSCUSSION**
Using DFT calculations embedded in ATK the tunnelling magneto-resistance value calculated for Fe/MgO/Fe magnetic tunnel junctions is found to be much greater than 1000, it is 2593.52%. However in case of hydrogen diffused (in its barrier) magnetic tunnel junction very poor conductance is shown in parallel orientation resulting in a much lowered value of TMR even below 500, precisely 467.96%. The reason for trapping of hydrogen atom is that during the fabrication process of any MTJ there is a chance of hydrogen atom to be trapped in any interstitial void of tunneling barrier because of its small size and favorable conditions. It becomes difficult for hydrogen to be trapped inside pure magnesium layer at relatively lower temperature but in most of fabrication processes temperature is much higher and in such conditions magnesium oxide layer with metal interface enhances probability of hydrogen diffusion into the voids[13-14].

For spin dependent tunneling transport conductance formula given by Landauer is :

$$G = \frac{e^2}{h} \sum_{k_{\parallel}} T(k_{\parallel}) \tag{1}$$

Where in equation (1) $\frac{e^2}{h}$ is known as conductance quantum and $T(k_{\parallel})$ is transmission probability as a function of transmission eigenvalues. Conduction channels perpendicular to current flow can be described as wave vectors with components parallel to electrode barrier interface. Solution of Schrodinger equation describing an electron as a wave is termed as wave function. $\psi_k(r) = V^{1/2} \exp(ik.r)$ is a wave function describing an electron wave as a solution to Schrodinger equation at a point 'r' in cell having volume V. An electron wave decaying completely away from electrode barrier interface is an evanescent state or wave.

$\Delta_1, \Delta_2, \Delta'_2$ and $\Delta_5$ are four symmetries possessed by any material, differentiated on the basis of atomic orbital sharing providing different decay rates to evanescent states. $\Delta_1$ state symmetry is possessed by MgO as there is sharing of 's' character of Mg with '$p_z$' character of nearest oxygen atom located at $\pm a/2$ distance[8-12]. $\Delta_1$ state symmetry provides slowest decay rate for evanescent electron wave which results in higher conductance value. $\Delta_1$ state symmetry can be observed as a circular pattern of brighter spots on contour plot of Spin dependent transmission spectrum with a function of in plane $k_{\parallel} = (k_A, k_B)$ wave vector. In Fig.3: (a), (c) a circular pattern can be seen followed by brighter region depicting symmetry followed by Fe/MgO/Fe MTJ. However, when hydrogen diffuses in MgO, it tends to make a bond with the oxygen atom and because of that orbital sharing of magnesium and oxygen is disturbed and symmetry gets distorted as clearly evident from Fig.4: (a), (c). Hence conduction through barrier is reduced in parallel orientation specifically and TMR gets reduced to a much lower value[12,15,18].

## CONCLUSION

In this work we have analysed the deterimental effect on conductance caused by diffusion of hydrogen atoms in interstitial voids during fabrication process of magnesium oxide barrier of an Fe/MgO/Fe magnetic tunnel junction, using first principle calculations. Higher value of TMR is favourable for read and write mechanism in storage devices and also improves the speed of Magnetic Random Access Memory. So fabrication of magnetic tunnel junction with better TMR values is required and this is possible by introducing improved oxidation methods and layer deposition techniques. Annealing of device should be carried out in very careful manner by keeping effective atmospheric parameters such as temperature and pressure during the process.

## ACKNOWLEDGEMENT

One of the authors, B.G., would like to thank the Department of Science and Technology of the government of India for partially funding this work.